\documentclass[11pt]{article}

\usepackage{epsfig}

\def\degs{\ifmmode ^{\circ}\else$^{\circ}$\fi}
\def\fdg{\hbox{$.\!\!^\circ$}}          

\begin{document}


\begin{center}
{\Huge The polarization evolution 
of the optical afterglow of GRB 030329}
\end{center}
\bigskip

\noindent
Jochen Greiner$^*$, Sylvio Klose$^\dag$, Klaus Reinsch$^\ddag$, Hans
Martin Schmid$^\ell$, Re'em Sari$^\$$, Dieter H. Hartmann$^\P$,
Chryssa Kouveliotou$^\S$, Arne Rau$^*$, Eliana Palazzi$^\diamondsuit$,
Christian Straubmeier$^{**}$, Bringfried Stecklum$^\dag$, Sergej
Zharikov$^{\dag\dag}$, Gaghik Tovmassian$^{\dag\dag}$, Otto
B\"arnbantner$^\sharp$, Christoph Ries$^\sharp$, Emmanuel
Jehin$^{\P\P}$, Arne Henden$^{\S\S}$, Anlaug A. Kaas$^{\ddag\ddag}$,
Tommy Grav$^{\$\$}$, Jens Hjorth$^{\ell\ell}$, Holger
Pedersen$^{\ell\ell}$, Ralph A.M.J. Wijers$^\#$, A. Kaufer$^{\P\P}$,
Hye-Sook Park$^{\diamondsuit\diamondsuit}$, Grant
Williams$^{\sharp\sharp}$, Olaf Reimer$^{\#\#}$

\bigskip
\bigskip

\noindent$^*$Max-Planck-Institut f\"ur extraterrestrische Physik, 85741 
Garching, Germany  

\noindent$^\dag$Th\"uringer Landessternwarte, 07778 Tautenburg, Germany

\noindent$^\ddag$Universit\"ats-Sternwarte G\"ottingen, 37083 G\"ottingen, 
Germany

\noindent$^\ell$Institut f\"ur Astronomie, ETH Z\"urich, 8092 Z\"urich, 
Switzerland

\noindent$^\$$California Institute of Technology, Theoretical Astrophysics 
130-33, Pasadena, CA 91125, USA

\noindent$^\P$Clemson University, Department of Physics and Astronomy, Clemson,
SC 29634, USA

\noindent$^\S$NSSTC, SD-50, 320 Sparkman Drive, Huntsville, AL 35805, USA

\noindent$^\diamondsuit$Istituto di Astrofisica Spaziale e Fisica
Cosmica, CNR, Sezione di Bologna, 40129 Bologna, Italy

\noindent$^{**}$ I. Physikalisches Institut, Universit\"at K\"oln, 50937 
K\"oln, Germany  

\noindent$^{\dag\dag}$Instituto de Astronomia, UNAM, 22860 Ensenada, Mexico

\noindent$^\sharp$Wendelstein-Observatorium, Universit\"atssternwarte, 
81679 M\"unchen, Germany

\noindent$^{\P\P}$European Southern Observatory, Alonso de Cordova 3107, 
Vitacura, Casilla 19001, Santiago 19, Chile

\noindent$^{\S\S}$Universities Space Research Association, U.S. Naval
Observatory, P.O. Box 1149, Flagstaff, AZ 86002, USA

\noindent$^{\ddag\ddag}$Nordic Optical Telescope, 38700 Santa Cruz de La Palma,
Spain

\noindent$^{\$\$}$University of Oslo, Institute for Theoretical Astrophysics, 
0315 Oslo, Norway, and Harvard-Smith\-sonian Center for Astrophysics, 
Cambridge, MA 02138, USA

\noindent$^{\ell\ell}$Astronomical Observatory, NBIfAFG, University of 
Copenhagen, 2100
Copenhagen, Denmark

\noindent$^\#$Astronomical Institute Anton Pannekoek, Kruislaan 403, 1098 
SJ Amsterdam, The
Netherlands

\noindent$^{\diamondsuit\diamondsuit}$Lawrence Livermore National
Laboratory, University of California, P.O. Box 808, Livermore, CA 94551, USA 

\noindent$^{\sharp\sharp}$MMT Observatory, University of Arizona, Tucson, 
 AZ 85721, USA

\noindent$^{\#\#}$Theoretische Weltraum-und Astrophysik, Ruhr-Universit\"at 
Bochum, 44780 Bochum, Germany

\bigskip
{\bf Abstract:} 
We report 31 polarimetric observations of the afterglow of GRB 030329 
with high signal-to-noise and high sampling frequency.
The data imply that the afterglow magnetic field 
has small coherence length and is mostly random, probably generated 
by turbulence.




\section{Introduction}

The association of a supernova with GRB 030329$^{1,2}$ strongly
supports the collapsar model$^3$ of $\gamma$-ray bursts (GRBs), where
a relativistic jet$^4$ forms after the progenitor star collapses. Such
jets cannot be spatially resolved because of their cosmological
distances. Their existence is conjectured based on breaks in GRB
afterglow light curves and the theoretical desire to reduce the GRB
energy requirements. Temporal evolution of polarization$^{5,6,7}$ may
provide independent evidence for the jet structure of the relativistic
outflow. 
Previous single measurements found low-level
(1-3\,\%) polarization$^{8-15}$ in optical afterglows, and the only
reports on variable polarization$^{16,17}$ were based on few
measurements with different instruments and modest signal-to-noise.
Here, we report polarimetric
observations of the afterglow of GRB 030329 with high signal-to-noise
and high sampling frequency$^{18}$. 

\section{Observations and Results}

GRB 030329 triggered the High Energy Transient Explorer, HETE-2, on
March 29, 2003 (11:37:14.67 UT)$^{19}$. The discovery of the burst
optical afterglow$^{20,21}$ was quickly followed by a redshift
measure\-ment$^{22}$ for the burster of z=0.1685 ($\sim$800 Mpc).
We have obtained 31 polarimetric
observations of the afterglow of GRB 030329 with the same
instrumentation (plus few more with different instruments)$^{18}$ over a time
period of 38\,days.
We performed relative photometry, and derived from each pair of
simultaneous measurements at orthogonal angles the Stokes parameters U
and Q. In order to obtain the intrinsic polarization of the GRB
afterglow, we had to correct for Galactic interstellar polarization
(mostly due to dust). We performed imaging polarimetry to derive the
polarization parameters of seven stars in the field of GRB 030329, and
obtained an interstellar (dust) polarization correction of 0.45\% at
position angle 155\degs. Subtraction of the mean foreground
polarization was performed in the Q/U plane
(Q$_{fp}$=0.0027$\pm$0.0013, U$_{fp}$=-0.0033$\pm$0.0017).  

The temporal evolution of the degree and angle of polarization
together with the R band photometry is shown in Figure 1,
demonstrating the presence of non-zero polarization,
$\Pi\sim0.3-2.5$\% throughout a 38-day period, with significant
variability in degree and angle on time scales down to hours. Further,
the spectropolarimetric data of the first three nights as well as the
simultaneous R and K band imaging polarimetry during the second night
show that the relative polarization and the position angle are
wavelength independent (within the measurement errors of about
0.1\,\%) over the entire spectral range. These data imply that
polarization due to dust in the host galaxy of GRB 030329 does not
exceed $\sim$0.3\%. 

Figure 1 shows that while the polarization properties show substantial
variability (for which no simple empirical relationship is apparent),
the R band flux is a sequence of power laws. During each of the power
law decay phases the polarization is of order few percent, different
from phase to phase, and variable within the phase, but not in tandem
with the ``bumps and wiggles'' in the light curve. We observe a
decreasing polarization degree shortly after the light curve break at
$\sim$0.4\,days (as determined from optical$^{21,23}$ and X-ray
data$^{24,25}$). Rapid variations of polarization occur
$\sim$1.5\,days after the burst, and could be related to the end of
the transition period towards a new power law phase starting at
$\sim$1.7\,days. Polarization eventually rises to a level of
$\sim$2\,\%, which remains roughly constant for another two weeks. 



\begin{figure}
  \includegraphics[height=.92\textwidth, angle=270]{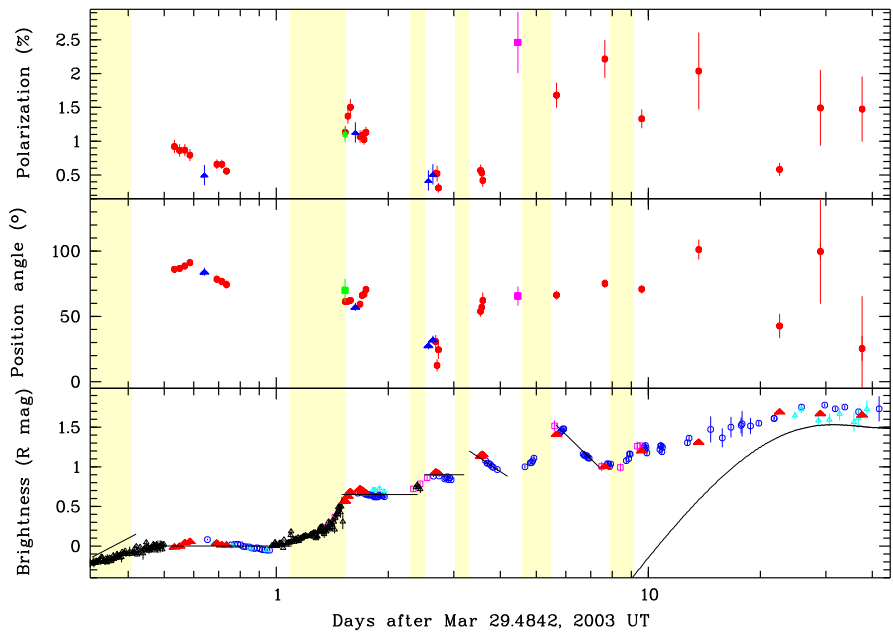}
  \caption{Evolution of the polarization during the first 38 days. Top
and middle panels show the polarization degree in percent and the
position angle in degrees. 
Spectropolarimetry 
was performed during the first three nights. 
The bottom panel shows the
residual R band light curve after subtraction of the contribution of a
power law t$^{-1.64}$ describing the undisturbed decay during the time
interval 0.5-1.2 days after the GRB (i.e., after the early break at
0.4 days), thus leading to a horizontal curve. The symbols correspond
to data obtained from either the literature (black), or own observations 
(gray): the 1m USNO
telescope at Flagstaff (circles), the OAN Mexico (open triangles), and 
FORS1/VLT (filled triangles). 
Lines indicate phases of power law decay, with the first one
from early data$^{21}$ (not shown). Vertical gray bars mark re-brightening
transitions. Contributions from an underlying supernova (solid curved
line) do not become significant until $\sim$10 days after the GRB.}
\end{figure}

Due to relativistic beaming, most of the observed photons arrive from a narrow 
cone, of opening angle $\theta$=1/$\Gamma$ around the line of sight. If the 
magnetic field parallel to the shock front is stronger than the perpendicular 
component, a resolved afterglow would look like a ring with polarization 
pointing towards its center$^5$. Early on, however, the Lorentz factor is 
much larger than the inverse opening angle of the jet (1/$\Gamma$), and we 
therefore probe scales that are much narrower than the size of the jet. 
The jet is uniform over these scales, so the emission pattern (afterglow) 
has axial symmetry around the line of sight. This symmetry, for an unresolved 
source, leads to zero polarization.  As the ejecta interact with the 
surrounding medium and decelerate, their Lorentz factor becomes comparable 
to 1/$\Gamma$, so we see most of the emitted photons. As the jet 
expands laterally, the energy per unit solid angle decreases; the light 
curve power law decay changes to a steeper slope, marking the so-called 
"jet break" and, at the same time, axial symmetry is broken (if the jet 
is not exactly pointed towards us), resulting in non-zero polarization. 
Since the jet spreads but the offset of the line of sight from the 
centre of the jet remains constant, axial symmetry is regained and 
polarization will eventually vanish. Maximal polarization should therefore 
occur around the jet break time. Some models$^{5,6,7}$ suggest the presence 
of either one or three peaks of polarization,  with the most significant 
peak close to the time of the jet break. If three peaks are 
present, the position angle of the central peak is rotated by 90\degs\ 
relative to the other two, but remains constant within each peak.

How do our data reflect these properties? A break (change of the power law 
decay in the light curve) was found at ~0.4 days in early optical 
data$^{21,23}$ and confirmed with X-ray data$^{24,25}$. We thus expect 
maximum polarization near 0.4 days, and a decline 
thereafter. Our early polarization data (Figure 1) are consistent with 
this prediction. Though we are missing data before 0.4 days to confirm a 
peak, the degree of polarization decreases from 0.9\% at ~0.55 days to 
about 0.5\% at 0.8 days. During this decline the position angle decreases
slightly, while the model predicts constancy. 
The time evolution of the polarization properties during this early phase 
is thus broadly consistent with the interpretation 
of the steepening of the light curve at 0.4 days as the 
jet break, providing independent observational evidence for the 
crucial assumption of collimated outflows (jets) in GRB explosions.

GRB 030329 is so far the only case where the polarization evolution supports 
the break as being due to the jet nature. From the jet break time and 
the isotropic equivalent energy of 9x10$^{51}$ erg we calculate the jet 
opening angle$^{26}$ to be about 3\fdg5, 
and the actual total energy release during the burst ~2x10$^{49}$ erg. 
This energy is about 25 times smaller than the "standard energy" of 
GRBs$^{27}$ and one order of magnitude larger 
than the one inferred (assuming isotropic emission) for GRB 980425 
(associated with SN 1998bw).

We do not expect the model to apply to the observations after the first 
re-brightening episode, which started $\sim$1.5 days after 
the trigger. Indeed the polarization angle changed by 30\degs\ with respect 
to the first night, while the model predicts either no change or a 90\degs\ 
change. What can we still learn from the complex late 
time behavior? Between 3 and 10 days the magnitude of the polarization changes 
significantly (a factor of two or more), but the position angle remains fairly
constant (fluctuations of less than 10 degrees). This implies that the 
polarization is not the result of a small number of coherent magnetic field 
cells with random orientation; such a model would predict that the position
angle changes on the same timescale as the magnitude of polarization. 
Instead, it implies that the position angle is associated with 
some global geometry.

A change of the position angle by an amount different than ninety degrees, as 
observed between the first and second day as well as between the second and 
third day, suggests that the asymmetry of the emission changed direction. 
Independent of polarization, the steeper decay of the optical light curve 
after day 10 (once the supernova component is subtracted), as well as a 
clear break observed at radio frequencies, suggests that the outflow may 
consist of two components$^{28}$: the first one 
dominates the light curve and the polarization properties until day 1.5, 
and has its jet break at day 0.4, and the second one, more mildly relativistic 
(Lorentz factor of about six), causes the re-brightening at 1.5 days and 
dominates the light curve and polarization properties thereafter. In this 
interpretation, the polarization data suggest that these two components do 
not share the same symmetry axis, which allows for the 30\degs\ 
change in the position angle. Due to its larger angle but similar energy 
per unit solid angle, the energy content in the second component is 
comparable to the canonical value of 5x10$^{50}$ ergs. 
Even  more than two jet components have been proposed$^{29}$ to 
explain the various wiggles in the lightcurve. If that picture is correct, 
the polarization data require that the jet axes not be aligned.

Finally, after day $\sim$10, there is a hint of some decrease in the 
polarization degree.
At this time, 
the supernova contributes approximately 60\% of the total 
light in the R band. Radiation transport models of non-symmetric 
supernovae$^{30,31}$ suggest a $\sim$1\% degree of polarization. 
For SN 2003dh we expect an even lower level of polarization since 
we are oriented towards the SN rotation axis. 
Thus, it is plausible that the 
low polarization supernova light dilutes the more highly polarized afterglow.

In summary, our data constitute the most complete and dense sampling
of the polarization behaviour of a GRB afterglow to date. 
The GRB 030329 afterglow polarization probably did not
rise above $\sim$2.5\%, and  did
not correlate with the flux. The low level of polarization implies
that the components of the magnetic field parallel and perpendicular
to the shock do not differ by more than $\sim$10\,\%, and suggests an
entangled magnetic field, probably amplified by turbulence behind
shocks, rather than a pre-existing field. 

\smallskip
\noindent{\bf Acknowledgements:}
This work is primarily based on observations collected at ESO, Chile,
with additional data obtained at the German-Spanish Astronomical
Centre Calar Alto, operated by the Max-Planck-Institute for Astronomy,
Heidelberg, jointly with the Spanish National Commission for
Astronomy, the NOT on La Palma, Canary Islands, and the Observatorio
Astronomico National, San Pedro, Mexico. We are grateful to the staff
at the Paranal, Calar Alto and NOT observatories, in particular
A. Aguirre, M. Alises, S. Hubrig, A.O. Jaunsen, C. Ledoux, S. Pedraz,
T. Szeifert, L. Vanzi and P. Vreeswijk for obtaining the service mode
data reported here.



\end{document}